# STM imaging of symmetry-breaking structural distortion
# in the Bi-based cuprate superconductors

Ilija Zeljkovic[1], Elizabeth J. Main[1], Tess L. Williams[1], M. C. Boyer[2], Kamalesh Chatterjee[2], W. D. Wise[2], Yi Yin[1], Martin Zech[1], Adam Pivonka[1], Takeshi Kondo[2,3], T. Takeuchi[3], Hiroshi Ikuta[3], Jinsheng Wen[4], Zhijun Xu[4], G. D. Gu[4], E. W. Hudson[1,2], Jennifer E. Hoffman[1,*]

[1] *Department of Physics, Harvard University, Cambridge, MA 02138, U.S.A.*

[2] *Department of Physics, Massachusetts Institute of Technology, Cambridge, MA 02139, U.S.A.*

[3] *Department of Crystalline Materials Science, Nagoya University, Nagoya 464-8603, Japan*

[4] *Condensed Matter Physics and Materials Sciences Department, Brookhaven National Laboratory, Upton, New York 11973, U.S.A.*

[*] *To whom correspondence should be addressed: jhoffman@physics.harvard.edu*

**A complicating factor in unraveling the theory of high-temperature (high-$T_c$) superconductivity is the presence of a "pseudogap" in the density of states, whose origin has been debated since its discovery[1]. Some believe the pseudogap is a broken symmetry state distinct from superconductivity[2-4], while others believe it arises from short-range correlations without symmetry breaking[5,6]. A number of broken symmetries have been imaged and identified with the pseudogap state[7,8], but it remains crucial to disentangle any electronic symmetry breaking from pre-existing structural symmetry of the crystal. We use scanning tunneling microscopy (STM) to observe an orthorhombic structural distortion across the cuprate superconducting $Bi_2Sr_2Ca_{n-1}Cu_nO_{2n+4+x}$ (BSCCO) family tree, which breaks two-dimensional inversion symmetry in the surface BiO layer. Although this inversion symmetry breaking structure can impact electronic measurements, we show from its insensitivity to temperature, magnetic field, and doping, that it cannot be the long-sought pseudogap state. To detect this picometer-scale variation in lattice structure, we have implemented a new algorithm which will serve as a powerful tool in the search for broken symmetry electronic states in cuprates, as well as in other materials.**

Numerous symmetry-breaking electronic states have been theoretically proposed to explain the cuprate pseudogap. A two-dimensional "checkerboard" charge density wave (CDW), may break translational but not rotational symmetry[9], while coexisting spin and charge density wave "stripes"[10] break both (though a precursor nematic state may break only rotational symmetry[11]). More exotic states, like the $d$-density wave, similarly break translational symmetry, but also time-reversal symmetry, while preserving their product[12]. Intra-unit-cell orbital current loops break time-reversal and inversion symmetry but also preserve their product[13]. The experimental realization of these symmetry breaking states, and their identification with the pseudogap, remain highly debated.

Strong electronic distortions are typically accompanied by structural distortions, and vice versa. Determining the relationship between these orders can be complicated, but a clue comes from their (co-)dependence on other parameters. In a cuprate superconductor, for example, both superconductivity and the pseudogap are highly dependent on doping, temperature, and magnetic field. Here, we investigate whether the structural symmetry in BSCCO is similarly dependent on these parameters, or whether it is an omnipresent background within which the electronic states evolve.

Structural symmetries are traditionally measured by scattering experiments, such as x-ray or neutron scattering to determine bulk symmetries, or low energy electron diffraction (LEED) to determine surface symmetries. The structure of double layer $Bi_2Sr_2CaCu_2O_{8+x}$ (Bi-2212) is sketched in Fig. 1a. Although nearly tetragonal, a ~0.5% difference between $a$ and $b$ axes[14] makes the true structure orthorhombic. However, despite numerous scattering experiments on BSCCO spanning two decades, the more detailed structure has remained enigmatic, due in part to an incommensurate structural "supermodulation" which pervades the bulk of these materials[14], and to dopant disorder which leads Bi atoms to stochastically occupy inequivalent sites in different unit cells[15].

Electronic states in Bi-based cuprates have been heavily investigated by a variety of techniques. Angle-resolved photoemission spectroscopy (ARPES) has shown electronic states breaking time-reversal symmetry[16], as well as particle-hole and translational symmetry[17]. STM has shown evidence for electronic "checkerboard" states breaking translational symmetry[2,18-21] and nematic states breaking rotational symmetry[7]. STM and transient grating spectroscopy (TGS) have also suggested that electronic states break local inversion symmetry[8]. Furthermore, STM has found nanoscale variations in these symmetry breaking states[18-23]. Thus, to investigate the role of structure in these broken symmetry electronic states, it is imperative to make atomic scale measurements of the structural symmetry.

To undertake this investigation, we use three different home-built scanning tunneling microscopes. In each case, a sample is cleaved at low temperature in cryogenic ultra-high vacuum, and immediately inserted into the scanning head. BSCCO typically cleaves between two BiO mirror planes (Fig. 1a). Data was acquired at T=6K unless otherwise noted. The tip is rastered across the sample surface, while a feedback loop adjusts its height to maintain a constant tip-sample tunneling current. This results in a topographic image of the BiO surface.

Temperature drift (typically < 10 mK), piezo hysteresis, and piezo nonlinearity, can lead to small but problematic warping of topographic images. Recently, Lawler *et al* introduced a ground-breaking algorithm to correct these picometer-scale drifts[7]. We show that Lawler's algorithm can also be used to remove subtle periodic noise (see Supplementary Information, section I). More importantly, Lawler's algorithm sets the stage for our new algorithm to create an ultra-high-resolution average unit cell or "supercell" by overlaying, with better than 10 picometer precision[24], portions of an STM topography separated by hundreds of unit cells (see Supplementary Information, section II). A supercell averaging technique was previously used for sub-unit-cell resolution of $Sr_2Ru_3O_7$ orbitals[25], but our algorithm doubles the spatial precision and allows fast, automatic averaging of more than 10,000 unit cells.

Using our new average supercell algorithm, we have investigated a variety of BSCCO systems. In figures 1b-d, we show three drift-corrected topographic images of optimally doped Bi-2212 with supermodulation, and of optimally doped $Bi_{2-y}Pb_ySr_2CuO_{6+x}$ (Bi-2201) where Pb doping has been used to completely, and partially, suppress the supermodulation. Insets in these topographies show average 2x2 supercell blocks. As expected, the incommensurate supermodulation has no discernible effect on the commensurate positions of the atoms within the average supercell. However, fits to determine the location of each atom within the supercell show consistent shifts of each atom by ~2% of the unit cell, in alternate directions along the orthorhombic *a* axis.

The global structural symmetry can also be seen in the Fourier transforms of these topographies (Figs. 1e-g). The dominant features are the tetragonal lattice vectors $Q_x$ and $Q_y$, and, in fields of view where the supermodulation has not been suppressed by Pb, the wavevector $Q_{sm}$. However, in each case we also find one, and only one, of the orthorhombic wavevectors $Q_a$ or $Q_b$, shown schematically in Fig. 2a. The presence of an orthorhombic wavevector confirms the in-plane relative shift of the two Bi atoms already noted in the average supercells (see Fig. 2b and Supplementary Information section III). This shift reduces the symmetry of the crystal by removing mirror planes perpendicular to the *b* and *c* axes and replacing them with glide planes, thus changing the space group of the crystal from orthorhombic Fmmm to orthorhombic Amaa. Crystals in this space group still have three-dimensional inversion symmetry; for instance in Bi-2212, the inversion center lies in the calcium layer. However, distortion removes the two-fold rotation symmetry along the *c* axis, effectively breaking two-dimensional inversion symmetry in the BiO plane.

The presence of only one of $Q_a$ or $Q_b$ indicates that, in Figs. 1b-d, the Bi shift is along the same orthorhombic lattice axis throughout the field of view. This Bi shift is consistent with a distortion observed in supermodulated and un-supermodulated single and double layer BSCCO, by multiple bulk techniques (see refs. 14 and 15, and additional references in table S1 of Supplementary Information, section IV). STM now becomes the first technique to measure this distortion locally.

The question remains: what relationship, if any, does this structural distortion bear to the variety of predicted and observed electronic orders, particularly to highly-debated claims[8,13] that the pseudogap is characterized by intra-unit-cell inversion symmetry breaking? To investigate this, we characterize the dependence of the structural distortion on parameters which are known to heavily influence electronic ordered states: doping, temperature, and magnetic field. Fig. 3a locates in a three-dimensional phase diagram the 21 datasets in which we measured the structure. The key results are summarized in Figs. 3b-d. We do not find a dependence of the structural distortion on doping, temperature, or field, across a wide range of values. We have measured the distortion both inside and outside the superconducting state, and in samples with vastly different pseudogap energies, yet the structural distortion appears insensitive to these effects. We therefore conclude that the structural inversion symmetry breaking state is an omnipresent background against which the electronic states evolve, and not the long-sought pseudogap state itself.

The supermodulation is another background structural phenomenon, producing strain which varies the $Cu-O_{apical}$ bond length by up to 12% with an incommensurate ~2.6 nm periodicity. The gap in the electronic density of states is also modulated with this periodicity, and it has been hypothesized that the strain is the primary cause[26]. This raises the question: does the strain of the orthorhombic distortion similarly locally modulate the pseudogap energy? In fact the pseudogap energy varies on the ~2-3 nm length scale[27], while the strain of the orthorhombic distortion flips direction at every other Bi atom (0.54 nm length scale). We further note that energy gaps from averaged spectra acquired at the two inequivalent Bi sites are indistinguishable (see Supplementary Information, section V). This demonstrates that the orthorhombic distortion strain does not modulate the pseudogap phase.

In non-Pb doped Bi-2212, we find that the orthorhombic distortion lies along the $a$-axis (wavevector at $Q_a$), orthogonal to the supermodulation in all ten samples investigated. The probability that this is a coincidence is less than 0.1%. In Pb-doped samples, however, the supermodulation is suppressed to the point where it no longer forces the orthorhombic distortion along the $a$-axis. Although without the supermodulation it is impossible for us to differentiate between the $a$ and $b$ axes, we did, as shown in Fig. 4, find one domain boundary, where the distortion rotates by 90°. This is rare – in all other images, up to 80 nm square, we find, as shown in Fig. 1, a single orthorhombic distortion vector. The probability of finding a domain boundary is on the order of $4L/D$, where $L$ is the size of our image and $D$ is the size of a domain; from 15 images of average size 50 nm, we estimate the size of a domain to be on the order of one micrometer.

To find domain boundaries we map the spatial dependence of the strength of the topographic signal $A$ at the orthorhombic wavevectors. This is calculated by demodulating the signal (shifting the Fourier transform so the desired wavevector $\vec{q}$ is at zero frequency, low-pass filtering and then inverse transforming), giving us a spatially dependent Fourier amplitude $A(\vec{q}, \vec{r})$. One of $A(\vec{Q}_a, \vec{r})$ and $A(\vec{Q}_b, \vec{r})$ will be zero and the other non-zero, so that a map of the difference, $D(\vec{r}) = A(\vec{Q}_a, \vec{r}) - A(\vec{Q}_b, \vec{r})$, will have a roughly uniform magnitude and sign. The spatial distributions of these maps are reflected in the error bars of Fig. 3. In Fig. 4b, however, the sign of $D(\vec{r})$ flips between the upper and lower half of the figures, highlighting the rotation of the distortion axis between these two regions. The difference is also clear in Fourier transforms from the two different regions (Figs. 4c-d).

In claiming a broken electronic symmetry state, one should know whether the electronic order breaks or preserves the structural symmetry of the crystal. Bulk probes such as x-ray diffraction and neutron scattering cannot determine whether electronic and structural orders choose the same symmetries locally; in contrast, STM can investigate these symmetries on atomic length scales. For example, we suggest that future STM studies should search for modification of the local electronic symmetry across structural domain boundaries, such as that shown in Fig. 4. Furthermore, as the structural inversion symmetry breaking may lead to the *appearance* of electronic inversion symmetry breaking, we suggest that any reports identifying this inversion symmetry breaking with the pseudogap phase should explicitly track the broken symmetry with temperature, magnetic field, and doping, as we have done, to ensure that it changes as would be expected with the changing pseudogap energy scale. We note that there is precedent for the misidentification of this structural effect as a long-sought electronic effect[28]. Finally, implementation of our average supercell algorithm, coupled with the ability of STM to measure structural and electronic symmetries locally, will become an important part of future STM studies, not only on cuprates, but on many important materials, such as the new iron-based superconductors, where the cleaved surface structure remains controversial[29]. Our real-space algorithm can even enable atom-specific high-resolution unit cell mapping not possible with Fourier techniques, in cases where two visibly different species are stochastically mixed at the sample surface, such as Se and Te site in $FeTe_{1-x}Se_x$ which can be sorted automatically based on their bimodal distribution of topographic height[30].

**Acknowledgements:** We are grateful for fruitful conversations with Erez Berg, Andrea Damascelli, Seamus Davis, Steve Kivelson, Dung-Hai Lee, Freek Massee, Joe Orenstein, Subir Sachdev, and Chandra Varma. This work is supported by the NSF, NSF Career, PECASE and Research Corporation. Work in BNL is supported by the US DOE. T.L.W. acknowledges support from an NDSEG fellowship.

**Author contributions:** STM measurements were done by M.C.B., K.C., W.D.W., Y.Y., M.Z., T.L.W., E.J.M, A.P. and I.Z. T.K., T.T., H.I., J.W., Z.X. and G.D.G. grew the samples. I.Z., E.J.M, and T.L.W. analyzed the data and wrote the paper with E.W.H. and J.E.H.

**Author Information:** The authors declare no competing financial interests. Correspondence and requests for materials should be addressed to J.E.H. (jhoffman@physics.harvard.edu).

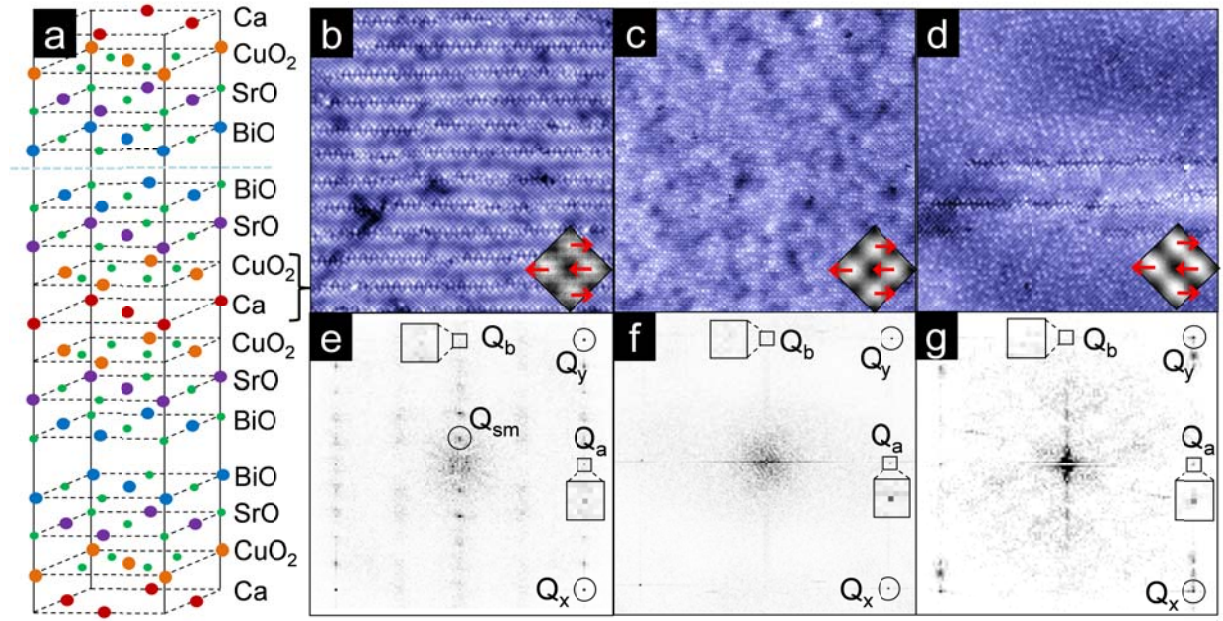

**Figure 1 | Crystal structure and surface topographies of Bi-based cuprates. a**, Schematic of $Bi_2Sr_2Ca_{n-1}Cu_nO_{2n+4+x}$, (n=2 as pictured, with the bracketed and similar layers removed or repeated for other n). Single crystals cleave between adjacent BiO layers (dashed line). **b-d**, Topographic images of 30 nm square regions of the BiO lattice from three different samples: **b**, optimally doped Bi-2212 with supermodulation; **c**, optimal $T_c = 35K$ Bi-2201 with Pb doping completely suppressing the supermodulation; and **d**, overdoped $T_c \sim 0$ K Bi-2201, also doped with Pb but with remnants of the supermodulation. Each was cropped from larger (up to 80nm) fields of view, for ease of viewing. Insets show the average supercell with arrows denoting the shifts of the Bi atoms. **e-g**, Fourier transforms of **b-d**, showing peaks at the tetragonal Bragg vectors $Q_x$ and $Q_y$ as well as at the orthorhombic $Q_a=(Q_x+Q_y)/2$ (but not at the equivalent $Q_b$). In **e** the supermodulation creates an additional incommensurate peak at $Q_{sm}$.

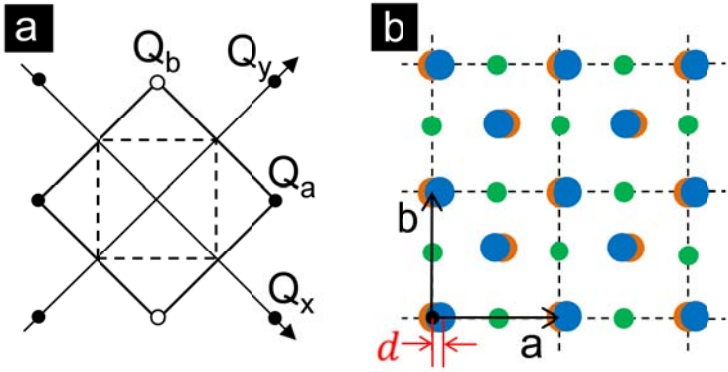

**Figure 2 | BiO lattice in real and momentum space. a**, In momentum space, Bragg vectors (black circles) at $Q_x=(2\pi/a_0, 0)$, $Q_y=(0, 2\pi/a_0)$, define the tetragonal Brillouin zone (black square). Orthorhombic vectors (circles) $Q_a=(\pi/a_0, \pi/a_0)$, $Q_b=(-\pi/a_0, \pi/a_0)$ define the orthorhombic Brillouin zone (dashed square). (The average Bi-Bi nearest neighbor distance is $a_0 = 0.383$ nm.) The orthorhombic distortion appears as a density wave along the a-axis, with alternate Bi rows moving closer/farther apart, leading to strength at $Q_a$ in the FTs. $Q_b$ (open circle) is notably absent. **b**, In real space, the distortion in the surface layer of Bi (blue) and O (green) atoms appears as shifts of two Bi sub-lattices in opposite directions along the orthorhombic a-axis. This distortion breaks inversion symmetry at the Cu site, but preserves a single mirror plane along the a-axis. The undistorted Cu lattice (orange), two layers beneath the Bi atoms, is shown for reference (but does not appear in topographic images). The displacement d, of one Bi atom from its undistorted position, is marked here and reported on the left axis of Figs. 3b-d as a fraction of the orthorhombic lattice constant a, and on the right axis of Figs. 3b-d in picometers.

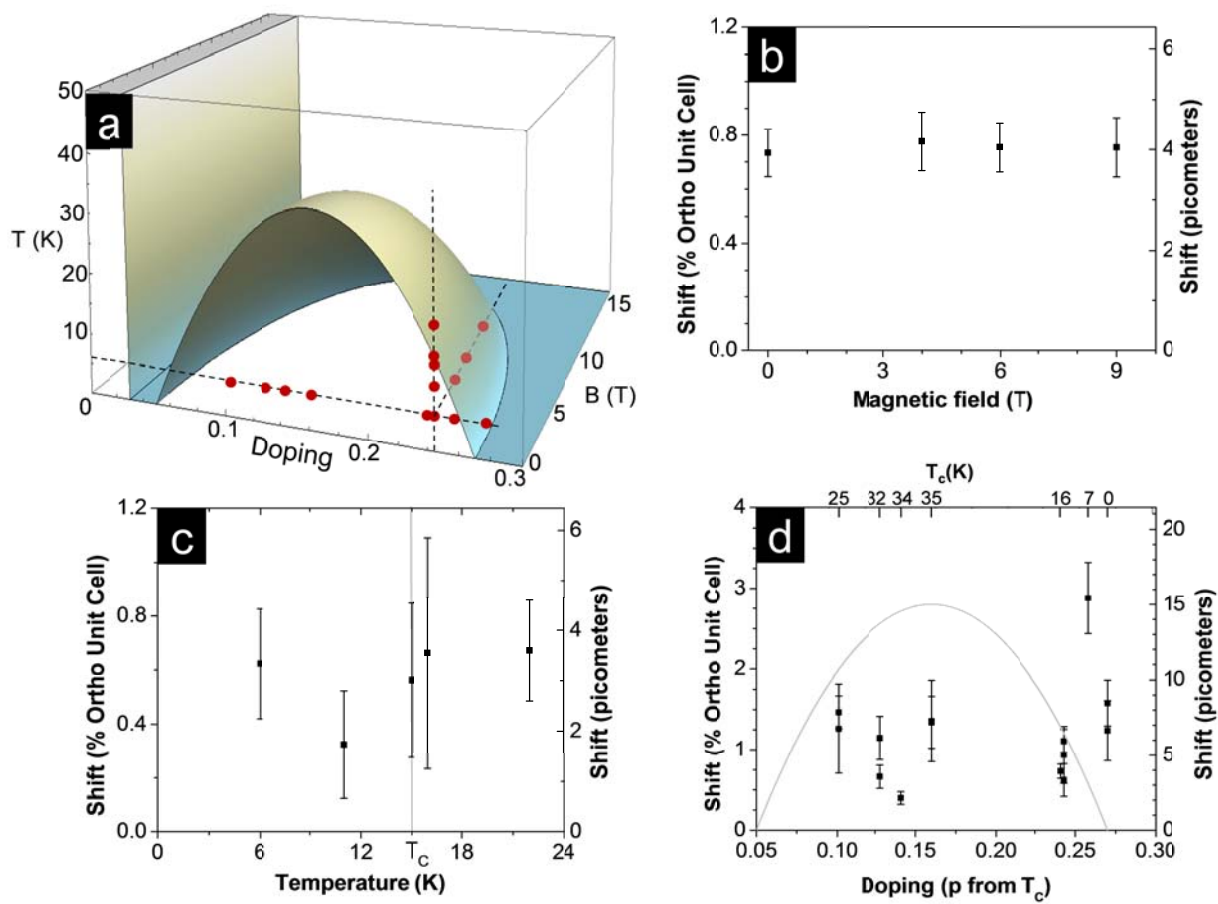

**Figure 3 | Orthorhombic distortion as a function of doping, temperature, and magnetic field**.
**a**, Schematic phase diagram of Bi-2201, locating (red dots) the 15 experimental conditions under which the 21 datasets used in this figure were taken (see details in Supplementary Information, section VI). **b-d** show orthorhombic distortion as a function of **b**, magnetic field; **c**, temperature; and **d**, doping[31]. **b** and **c** each contain data from a single tip and region of overdoped Bi-2201 samples with $T_c$ = 16 K and 15 K respectively. Additional scatter in **d** may reflect differences in tip quality or field of view across the 14 samples. However, in all cases a clear, non-zero distortion was found along only one of the two orthorhombic axes.

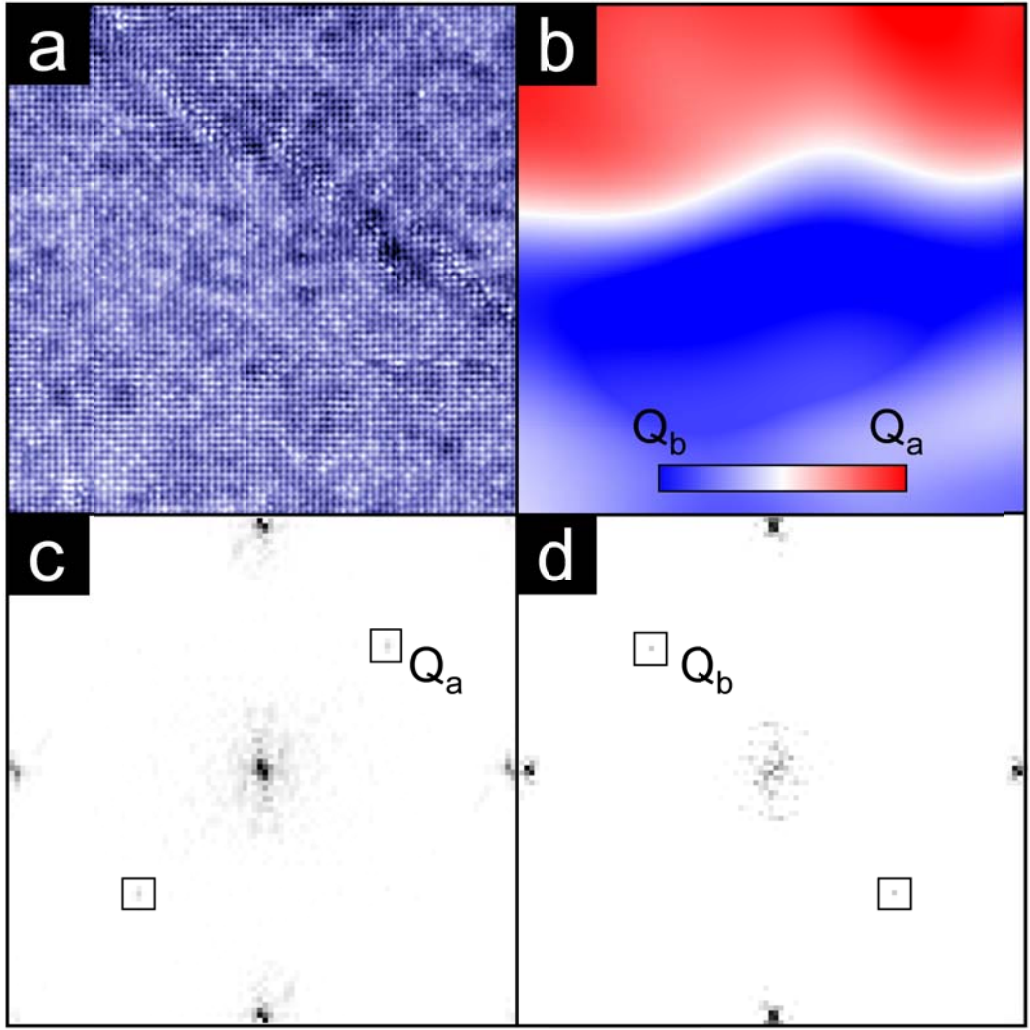

**Figure 4 | Orthorhombic domain boundary. a**, Topographic image of $T_c$ = 25K underdoped Bi-2201. **b**, The distortion map, $D(\vec{r}) = A(\vec{Q}_a, \vec{r}) - A(\vec{Q}_b, \vec{r})$, of the same field of view, where $A(\vec{q}, \vec{r})$ is the spatially dependent Fourier amplitude. A change in sign (color) indicates the rotation of the distortion axis and associated mirror plane across the domain boundary separating top and bottom of the figure. **c-d**, Fourier transforms of the top and bottom of **a** respectively, highlight the rotation of the orthorhombic distortion vector from $\vec{Q}_a$ to $\vec{Q}_b$.

## Supplementary Information

### (I) Differentiating between noise and true periodic structures in STM images

Two-dimensional Fourier transforms (FT) of STM topographic images of BSCCO reveal many peaks. In addition to the structural Bragg peaks which are universally observed, the supermodulation peaks which appear in non-Pb-doped samples, and the peaks related to charge ordering which have recently been reported[1] we notice several other peaks. It can be challenging to distinguish between meaningful peaks which represent true spatial periodicities within the sample, and peaks which arise from time-periodic environmental noise. Using Lawler's algorithm[2], we adjust for the picometer drift which occurs during a many hour map, and collapse the Bragg peaks onto a single pixel. Furthermore, any other peaks associated with modulation in the image commensurate with the lattice should also collapse to a single pixel (Fig. S1). Upon application of Lawler's algorithm, it can be seen that both the Bragg peaks and the orthorhombic distortion peak (circled in purple) do collapse onto a single pixel, whereas other peaks (circled in red) broaden. Therefore, we have introduced a new use of Lawler's algorithm to differentiate between the Fourier peaks intrinsic to the sample, and external noise. This technique, demonstrated here for topographic images, can be extended for the analysis of dI/dV maps which measure modulations in the electronic density of states.

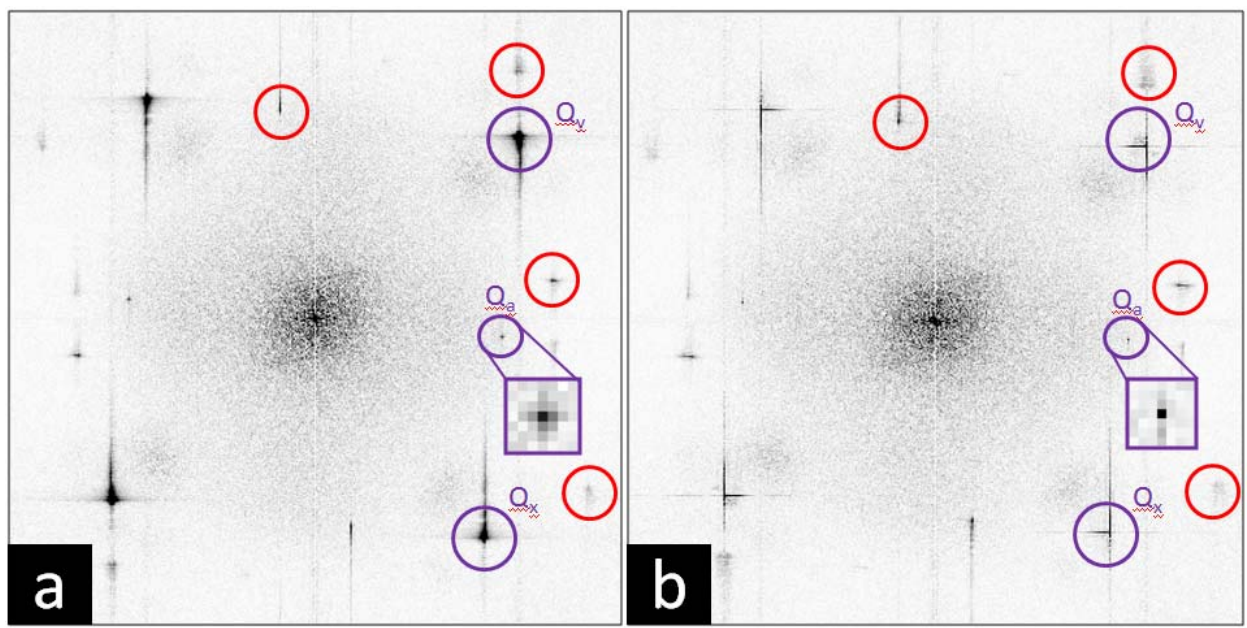

**Figure S1 | Drift correction of true signals vs. noise.** Fourier transform of a slightly underdoped Bi-2201 topography is shown **a**, before, and **b**, after application of Lawler's algorithm. After drift correction, Bragg peaks ($Q_x$ and $Q_y$) and orthorhombic distortion peak ($Q_a$) collapse onto a single pixel that is at least 5 times brighter than surrounding pixels . However, peaks circled in red broaden instead of collapsing, and are thus attributed to external noise.

## (II) Extraction of the average unit cell

Lawler's algorithm enables the determination of the true position of each individual atom in the lattice with picoscale precision. We extend this algorithm to calculate the average unit cell from the whole field of view. Fig. S2 demonstrates the use of the algorithm to create a 1 x 1 average unit cell.

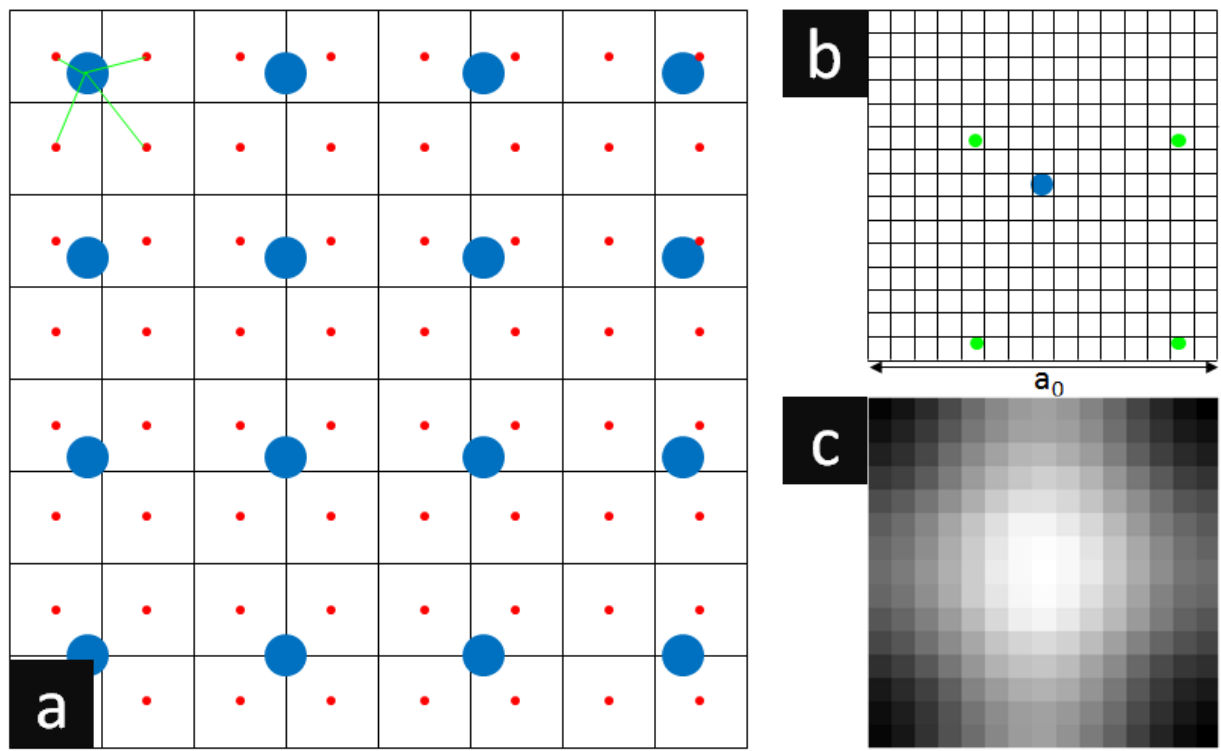

**Figure S2 | Illustration of average unit cell algorithm. a,** Schematic of a drift-corrected topographic image with Bi atoms represented as blue circles. The data is acquired in a pixel grid outlined in black; the center of each pixel is marked with a red dot. The resolution of this image is only slightly better than the Nyquist frequency for atoms. **b,** Schematic of the average 1 x 1 unit cell we will create with 15 x 15 pixels. We calculate the exact distance of every pixel in **a** to the nearest Bi atom (4 example distances are shown in green in **a**), then place it in the average unit cell in **b** to create a weighted histogram at each sub-unit-cell pixel. **c,** Example of the final average unit cell obtained from a real topographic image.

Starting with a drift-corrected topography (schematic shown in Fig. S2a), we create an average single-Bi unit cell with more pixels per atom than our raw data (typically 15 x 15 pixels per unit cell). For each pixel in the actual topography, the location can be calculated with picoscale precision in relation to the nearest Bi atom. The data from that pixel is then placed in the appropriate bin in the average unit cell (Fig. S2b). This process builds up a histogram of weight at each sub-unit-cell-resolved position in the average unit cell, ultimately showing a high-resolution map of an average Bi atom (Fig. S2c). An analogous process can be used to create an average supercell of any size. In the main text, we have calculated a 2 x 2 atom supercells in Fig. 1b-d insets.

The creation of an average supercell allows an improvement of intra-unit-cell spatial resolution, at the expense of inter-unit-cell variation. This method allows us to detect variations in atomic positions as small as a few thousandths of a unit cell. For example, from a 2 x 2 super-cell, we fit a peak to each of the four "atoms" in order to calculate their exact positions, and find orthorhombic distortion shifts as small as 0.5% of a unit cell (see values and error bars in Fig. 3, main text).

We compare our technique to the original supercell averaging technique used by Lee *et al*[3]. Lee's starting point for aligning the different unit cells to be averaged was a real-space Gaussian fit to each atomic peak in the topography. Their fitting procedure allowed location of the atom with 20 picometer accuracy, but proved sufficiently time consuming that only 28 unit cells were ultimately averaged. Our algorithm which combines Lawler's drift correction and supercell averaging can locate individual atoms in un-supermodulated BSCCO with typical ~8 picometer absolute accuracy and ~2 picometer relative accuracy within a single field of view[4]. Furthermore, the centers of these atoms can be automatically located, and the resultant average supercell computed, for >10,000 unit cells in just a few seconds of run time.

## (III) Calculation of the orthorhombic distortion by demodulating the Fourier component of the orthorhombic distortion wavevector

The orthorhombic distortion is measurable in both Fourier space (by observing the peak at $Q_a$ in the 2D Fourier transform), and real space (by creating a 2 x 2 super-cell and comparing the positions of the inequivalent atoms). We derive here the relationship between the two observed quantities: the value of the complex Fourier component at $\overrightarrow{Q_a}$ (shown in Fig. 2a), and the displacement $d$ (shown in red in Fig. 2b) directly measured in the average supercell.

We start by representing the topographic signal $T(\vec{r})$ and its spatial derivative $\nabla_{\vec{r}}\big(T(\vec{r})\big)$ in terms of the Bragg peak wavevectors $\overrightarrow{Q_x}$ and $\overrightarrow{Q_y}$, the wavevector of the orthorhombic distortion $\overrightarrow{Q_a}$, and the complex Fourier component $A$ at wavevector $\overrightarrow{Q_a}$. (We normalize the expression so that a Cu atom sits at the origin, and the strength of the signal at wavevectors $\overrightarrow{Q_x}$ and $\overrightarrow{Q_y}$ is equal to 1.)

$$T(\vec{r}) = e^{i\overrightarrow{Q_x}\cdot\vec{r}} + e^{-i\overrightarrow{Q_x}\cdot\vec{r}} + e^{i\overrightarrow{Q_y}\cdot\vec{r}} + e^{-i\overrightarrow{Q_y}\cdot\vec{r}} + Ae^{i\overrightarrow{Q_a}\cdot\vec{r}} + A^*e^{-i\overrightarrow{Q_a}\cdot\vec{r}} \tag{S1}$$

$$\nabla_{\vec{r}}(T(\vec{r})) = i\overrightarrow{Q_x}e^{i\overrightarrow{Q_x}\cdot\vec{r}} - i\overrightarrow{Q_x}e^{-i\overrightarrow{Q_x}\cdot\vec{r}} + i\overrightarrow{Q_y}e^{i\overrightarrow{Q_y}\cdot\vec{r}} - i\overrightarrow{Q_y}e^{-i\overrightarrow{Q_y}\cdot\vec{r}} + Ai\overrightarrow{Q_a}e^{i\overrightarrow{Q_a}\cdot\vec{r}} - A^*i\overrightarrow{Q_a}e^{-i\overrightarrow{Q_a}\cdot\vec{r}} \tag{S2}$$

Our objective is to find the value of $\vec{r}$ at which $T(\vec{r})$ reaches its maximum. This is the position in space to which the distorted Bi atom moves, with respect to an undistorted (0,0) position. Setting the gradient of $T(\vec{r})$ to 0, and using the fact that $\overrightarrow{Q_a} = \frac{\overrightarrow{Q_x + Q_y}}{2}$, we find

$$\overrightarrow{Q_x}(e^{i\overrightarrow{Q_x}\cdot\vec{r}} - e^{-i\overrightarrow{Q_x}\cdot\vec{r}} + \frac{A}{2}e^{i\overrightarrow{Q_a}\cdot\vec{r}} - \frac{A^*}{2}e^{-i\overrightarrow{Q_a}\cdot\vec{r}}) + \overrightarrow{Q_y}(e^{i\overrightarrow{Q_y}\cdot\vec{r}} - e^{-i\overrightarrow{Q_y}\cdot\vec{r}} + \frac{A}{2}e^{i\overrightarrow{Q_a}\cdot\vec{r}} - \frac{A^*}{2}e^{-i\overrightarrow{Q_a}\cdot\vec{r}}) = 0$$

$$(S3)$$

Since $\overrightarrow{Q_x}$ and $\overrightarrow{Q_y}$ are orthogonal, both parenthetic expressions must vanish. Because the parenthetic expressions are equal, we see that $\overrightarrow{Q_x}\cdot\vec{r} = \overrightarrow{Q_y}\cdot\vec{r}$. Because $\overrightarrow{Q_a} = \frac{\overrightarrow{Q_x} + \overrightarrow{Q_y}}{2}$, we see that $\overrightarrow{Q_a}\cdot\vec{r} = \overrightarrow{Q_x}\cdot\vec{r} = \overrightarrow{Q_y}\cdot\vec{r}$. In full generality, we express the complex distortion strength $A$ in terms of its half amplitude and phase, $A = 2A_0 e^{i\theta}$. Therefore,

$$\sin(\overrightarrow{Q_x}\cdot\vec{r}) + A_0\sin(\overrightarrow{Q_a}\cdot\vec{r} + \theta) = 0; \ \sin(\overrightarrow{Q_y}\cdot\vec{r}) + A_0\sin(\overrightarrow{Q_a}\cdot\vec{r} + \theta) = 0$$

$$\Rightarrow \sin(\vec{Q}_a\cdot\vec{r}) + A_0\sin(\vec{Q}_a\cdot\vec{r} + \theta) = 0$$

$$(S5)$$

From this expression, we find the distortion $d$ as a percent of the orthorhombic unit cell.

$$2\pi d = \overrightarrow{Q_a}\cdot\vec{r} = \frac{1}{2\pi}\arctan\left(\frac{-A_0\sin(\theta)}{1 + A_0\cos(\theta)}\right)$$

$$(S6)$$

One can simplify this expression considerably by noting that the orthorhombic distortion is odd about each Cu site. In other words, the ideal topography $T(\vec{r})$ could be expressed more simply as

$$T(\vec{r}) = 2\cos(\vec{Q}_x\cdot\vec{r}) + 2\cos(\vec{Q}_y\cdot\vec{r}) + 4A_0\sin(\vec{Q}_a\cdot\vec{r}) \qquad (S7)$$

(using the same normalization as in eqn S1). Therefore, for the perfect lattice, $\theta = -\pi/2$. In this case we find simply that $2\pi d = \arctan(A_0)$.

However, imperfections in the application of the drift correction algorithm, at most ~3% of a unit cell[4], may introduce a small phase error, so that $\theta = -\pi/2 + \varepsilon$. We then find

$$2\pi d = \arctan\left(\frac{A_0\cos\varepsilon}{1 + A_0\sin\varepsilon}\right)$$

$$(S8)$$

Because all quantities ($d$, $A_0$, and $\varepsilon$) are small, we expand the argument as

$$2\pi d \approx \arctan\left(A_0\left(1 - \tfrac{1}{2}\varepsilon^2\right)(1 - A_0\varepsilon)\right) \approx \arctan\left(A_0\left(1 - A_0\varepsilon - \tfrac{1}{2}\varepsilon^2\right)\right) \approx \arctan(A_0) \qquad (S9)$$

In conclusion, the complex value $A$ of the Fourier transform at the $\vec{Q}_a$ point is expected to be purely imaginary. Small (up to 3% unit cell) errors from imperfect drift correction may introduce a small real component to $A$, but such errors contribute only a second order correction to the relation $2\pi d = \arctan(A_0)$.

## (IV) Reported values of orthorhombic distortion in the $Bi_2Sr_2Ca_{n-1}Cu_nO_{2n+4+x}$ family tree

Table S1 shows the values of orthorhombic distortions reported in literature on single crystal $Bi_2Sr_2Ca_{n-1}Cu_nO_{2n+4+x}$ family of materials. ARPES, LEED, and TEM also report seeing the orthorhombic distortion in the BiO layer[5,6], but have not quantified the distortion. Using an STM, we do not have direct access to the distortion in the Cu-oxide layer. However, the average distortion we measure in the BiO layer (approximately 1% of the orthorhombic unit cell for Bi-2201 and 1.5% for Bi-2212) is comparable to the reported values using other techniques. Some crystallographers report the shift of Bi atoms along both axes in the BiO layer, but this is heavily dependent on the refining model used. In the case of centro-symmetric model in the orthorhombic space group Cccm (or less-conventional Amaa or Bbmb), a shift is observed along only one axis, whereas for non-centro-symmetric model in the lower symmetry orthorhombic space group A2aa (or A2/a), a shift is often concluded to be along both axes.

A significant advantage of STM is that we do not have to assume a model to extract the orthorhombic shift; we simply observe this shift in real space. In all 10 samples of supermodulated Bi-2212 we worked with, the shift is always orthogonal to the direction of the supermodulation, thus consistent with orthorhombic space group Amaa.

| Material | Expt. | Bi Distortion | Cu Distortion | Direction of Supermodulation | Structure | Reference |
|---|---|---|---|---|---|---|
| Bi-2212 | XRD | 2.22% **b** | -0.01% **b** | **a** | Amaa | Subramanian[7] |
| Bi-2212 | XRD | 2.15% **b** | 0.02% **b** | **a** | Amaa | Gao[8] |
| Bi-2201 | XRD | 2.58% **b** | none | **a** | Amaa | Torardi[9] |
| Bi-2201 | ND | 3.5% **b** | none | **a** | A2/a | Beskrovnyi[10] |
| Bi-2212 | ND | 2.55% **a** | -0.07% **a** | **b** | Bbmb | Miles[11] |
| Bi-2212 | XRD | 0.5% **a** – 1.73% **b** | -0.01% **b** | **a** | A2aa | Petricek[12] |
| Bi,Pb-2212 | XRD | -1% **a** +1.65% **b** | -2.57% **a** -0.02% **b** | none | Pnan | Calestani[13] |
| Bi,Pb-2212 | XRD | 1.1% **a** + 1.53% **b** | 0.08% **b** | none | A2aa | Gladyshevskii[14] |
| Bi,Pb-2201 | XRD | 1.82% **a** | none | none | Amaa | Ito[15] |
| Bi,Pb-2201 | STM | ~1% **a** | n/a | none | | present paper |
| Bi-2212 | STM | ~1.5% **a** | n/a | **b** | | present paper |

**Table S1 | Compilation of orthorhombic distortion literature.** All measurements listed here, except for the present result, are from bulk scattering experiments (x-ray diffraction and neutrons) using single crystals.

**(V)    Gap at the two inequivalent Bi sites**

To investigate the spectral effect of strain from the inversion symmetry breaking orthorhombic distortion, we compare the gap at the two inequivalent Bi sites in the orthorhombic unit cell. We find no evidence that the gap between these two sites. A representative example (chosen from our largest image) is shown in Fig. S3.

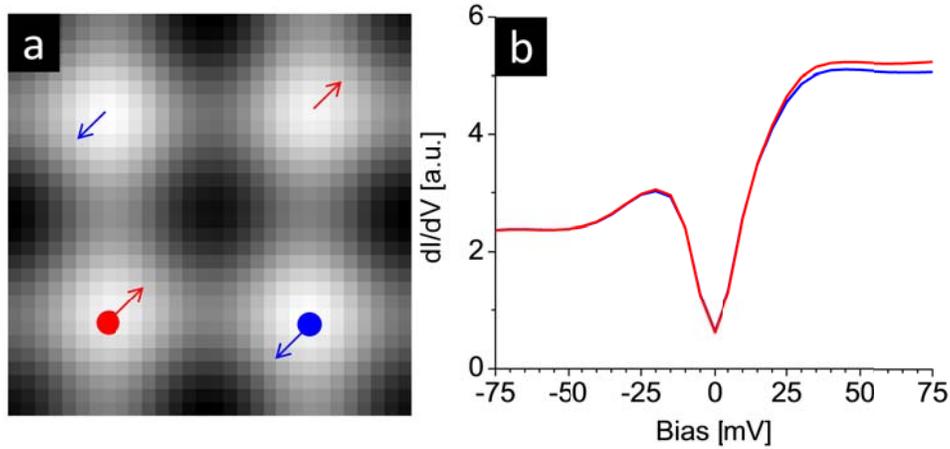

**Figure S3 | Gap at the two inequivalent Bi sites. a,** 2 x 2 average supercell from $T_c$=35 K underdoped Bi-2201 with Pb to suppress the supermodulation. **b,** Average spectra from the two inequivalent Bi sites show identical gap energy.

## (VI)   Experimental parameters for reported datasets

Three different home-built low-temperature STMs were used to acquire the data in this paper: one UHV STM in the Hudson laboratory at MIT, and two STMs in the Hoffman lab at Harvard, both with 9T magnetic field, one with cryogenic vacuum and the second with full system UHV. Table S2 shows STM setup conditions (current and bias), as well as temperature and magnetic field used for different samples we report on. Ten Bi-2212 datasets (not shown in table) were taken with the Hudson STM (7 samples) and the second Hoffman STM (3 samples).

| Figure | Material | Tc | T (K) | B (T) | Iset (pA) | Vset (mV) | Size (nm) | # pix | STM |
|--------|----------|------|-------|-------|-----------|-----------|-----------|-------|---------|
| 1b,e | Bi-2212 | OD80K | 6 | 0 | 30 | -100 | 30 | 232 | Hudson |
| 1c,f | Bi-2201 | OP35K | 6 | 0 | 400 | -100 | 30 | 194 | Hudson |
| 1d,g | Bi-2201 | OD0K | 6 | 0 | 50 | -250 | 30 | 234 | Hudson |
| 3b | Bi-2201 | OD16K | 6 | 0 | 100 | 100 | 40 | 256 | Hoffman |
| 3b | Bi-2201 | OD16K | 6 | 4 | 100 | 100 | 40 | 256 | Hoffman |
| 3b | Bi-2201 | OD16K | 6 | 6 | 100 | 100 | 40 | 256 | Hoffman |
| 3b | Bi-2201 | OD16K | 6 | 9 | 100 | 100 | 40 | 256 | Hoffman |
| 3c | Bi-2201 | OD15K | 6 | 0 | 100 | -100 | 18 | 122 | Hudson |
| 3c | Bi-2201 | OD15K | 11 | 0 | 100 | -100 | 18 | 122 | Hudson |
| 3c | Bi-2201 | OD15K | 15 | 0 | 100 | -100 | 18 | 122 | Hudson |
| 3c | Bi-2201 | OD15K | 16 | 0 | 100 | -100 | 18 | 122 | Hudson |
| 3c | Bi-2201 | OD15K | 22 | 0 | 100 | -100 | 18 | 122 | Hudson |
| 3d | Bi-2201 | OD0K | 6 | 0 | 150 | -250 | 17 | 256 | Hudson |
| 3d | Bi-2201 | OD0K | 6 | 0 | 50 | -250 | 58 | 450 | Hudson |
| 3d | Bi-2201 | OD7K | 6 | 0 | 100 | -100 | 26 | 450 | Hudson |
| 3d | Bi-2201 | OD15K | 6 | 0 | 100 | -100 | 18 | 122 | Hudson |
| 3d | Bi-2201 | OD15K | 6 | 0 | 100 | -100 | 44 | 485 | Hudson |
| 3d | Bi-2201 | OD 16K | 6 | 0 | 100 | 100 | 40 | 256 | Hudson |
| 3d | Bi-2201 | OP35K | 6 | 0 | 400 | -100 | 65 | 468 | Hudson |
| 3d | Bi-2201 | OP35K | 6 | 0 | 400 | -100 | 40 | 256 | Hudson |
| 3d | Bi-2201 | UD34K | 6 | 0 | 300 | -100 | 20 | 400 | Hudson |
| 3d | Bi-2201 | UD32K | 6 | 0 | 400 | -100 | 50 | 300 | Hudson |
| 3d | Bi-2201 | UD32K | 6 | 0 | 400 | -200 | 35 | 400 | Hudson |
| 3d | Bi-2201 | UD25K | 6 | 0 | 400 | -100 | 25 | 128 | Hudson |
| 3d | Bi-2201 | UD25K | 6 | 0 | 400 | -150 | 42 | 314 | Hudson |
| 4 | Bi-2201 | UD25K | 35 | 0 | 400 | -100 | 29.5 | 160 | Hudson |
| S3 | Bi-2201 | OP35K | 6 | 0 | 400 | -100 | 65 | 468 | Hudson |

**Table S2 | Experimental parameters for datasets used in main text.**

**Supplementary References**